\documentclass[twocolumn,preprintnumbers,amsmath,amssymb,superscriptaddress,longbibliography,nofootinbib,prl]{revtex4-1}
\usepackage{filecontents}
\usepackage{graphicx}
\usepackage{bm}
\usepackage{dsfont}
\usepackage[usenames,dvipsnames]{xcolor}
\usepackage{pstricks}
\usepackage[tight]{subfigure}
\usepackage{verbatim}
\usepackage{units}
\usepackage{multirow}
\usepackage{enumitem}
\usepackage{mathrsfs}
\usepackage{leftidx}
\usepackage{xspace}
\usepackage{braket}
\usepackage{bbm}
\usepackage{cancel}
\usepackage{amsfonts,amssymb,amsmath}
\usepackage[normalem]{ulem}

\usepackage{xr}
\usepackage{filecontents}

\usepackage[a4paper]{hyperref}
\hypersetup{colorlinks=true,linktoc=all,linkcolor=blue,breaklinks=true,citecolor=blue,urlcolor=blue}
\usepackage[english]{babel}
\usepackage{cleveref}

\renewcommand{\P}{\mathcal{P}}

\begin{document}

\title{Kinetic uncertainty relations for quantum transport}

\author{Didrik Palmqvist}
	\affiliation{Department of Microtechnology and Nanoscience (MC2), Chalmers University of Technology, S-412 96 G\"oteborg, Sweden}

\author{Ludovico Tesser}
	\affiliation{Department of Microtechnology and Nanoscience (MC2), Chalmers University of Technology, S-412 96 G\"oteborg, Sweden}

\author{Janine Splettstoesser}
	\affiliation{Department of Microtechnology and Nanoscience (MC2), Chalmers University of Technology, S-412 96 G\"oteborg, Sweden}
	
\date{\today}

\begin{abstract}
We analyze the precision of generic currents in a multi-terminal quantum-transport setting. Employing scattering theory, we show that the precision of such currents is limited by a function of the particle-current noise that can be interpreted as the activity in the classical limit. We thereby establish a kinetic uncertainty relation for quantum transport. In the full quantum limit, we find precision bounds with modified activity constraints depending on whether the system is fermionic or bosonic. We expect these bounds to be suitable as guideline for any transport process aiming at high precision.
\end{abstract}

\maketitle


Precision---namely the ratio between the square of average currents and their fluctuations---plays a key role in the performance of small-scale devices, e.g., when acting as thermodynamic machines. While in equilibrium~\cite{Callen1951Jul,Kubo1957Jun} or in specific nonequilibrium situations~\cite{Altaner2016Oct,Dechant2020Mar} fluctuation-dissipation theorems directly relate average quantities of interest to their fluctuations, such relations are more challenging to find under general nonequilibrium  conditions~\cite{Rogovin1974Jul,Levitov2004Sep,Safi2014Jan}.
Based on fluctuation relations~\cite{Esposito2009Dec,Campisi2011Jul,Harris2007Jul,Seifert2012Nov}, which hold even far from equilibrium but are often limited to classical Markovian or weak-coupling scenarios, a number of inequalities have been developed setting constraints on the achievable precision~\cite{Jarzynski2011Mar,Landi2024Apr}. 
These fundamental bounds include the thermodynamic uncertainty relation (TUR), constraining precision by the entropy production~\cite{Barato2015Apr,Horowitz2020Jan} {and being most predictive close to equilibrium}, and the kinetic uncertainty relation (KUR), constraining precision by the dynamical activity~\cite{DiTerlizzi2018Dec,Yan2019Sep,Hiura2021May} {and being most predictive far from equilibrium}, as well as combinations of those~\cite{Vo2022Sep,Hasegawa2024Feb}. Recently, there have been extensive efforts to extend the TUR~\cite{Guarnieri2019Oct} as well as the KUR~\cite{VanVu2022Apr,Hasegawa2023May,Prech2023Jun,Nishiyama2024Apr,Prech2024Jun,Bourgeois2024Jun} to the quantum regime, mostly exploiting Lindblad jump operators for the weak coupling limit. 

However, especially when dealing with energy-converting devices, where the goal is to generate currents of significant magnitude, strongly coupled systems under nonequilibrium conditions are of interest. In this strong-coupling limit, which has been studied in TURs~\cite{Brandner2018Mar,Liu2019Jun,Potanina2021Apr}, but also in recently developed fluctuation-dissipation bounds~\cite{Tesser2024May}, quantum statistics can play an important role.
Kinetic uncertainty relations for quantum transport~\cite{Prech2024Jul}, valid for strong coupling, are missing.

\begin{figure}[b!]
    \centering
   \includegraphics[width=3.3in]{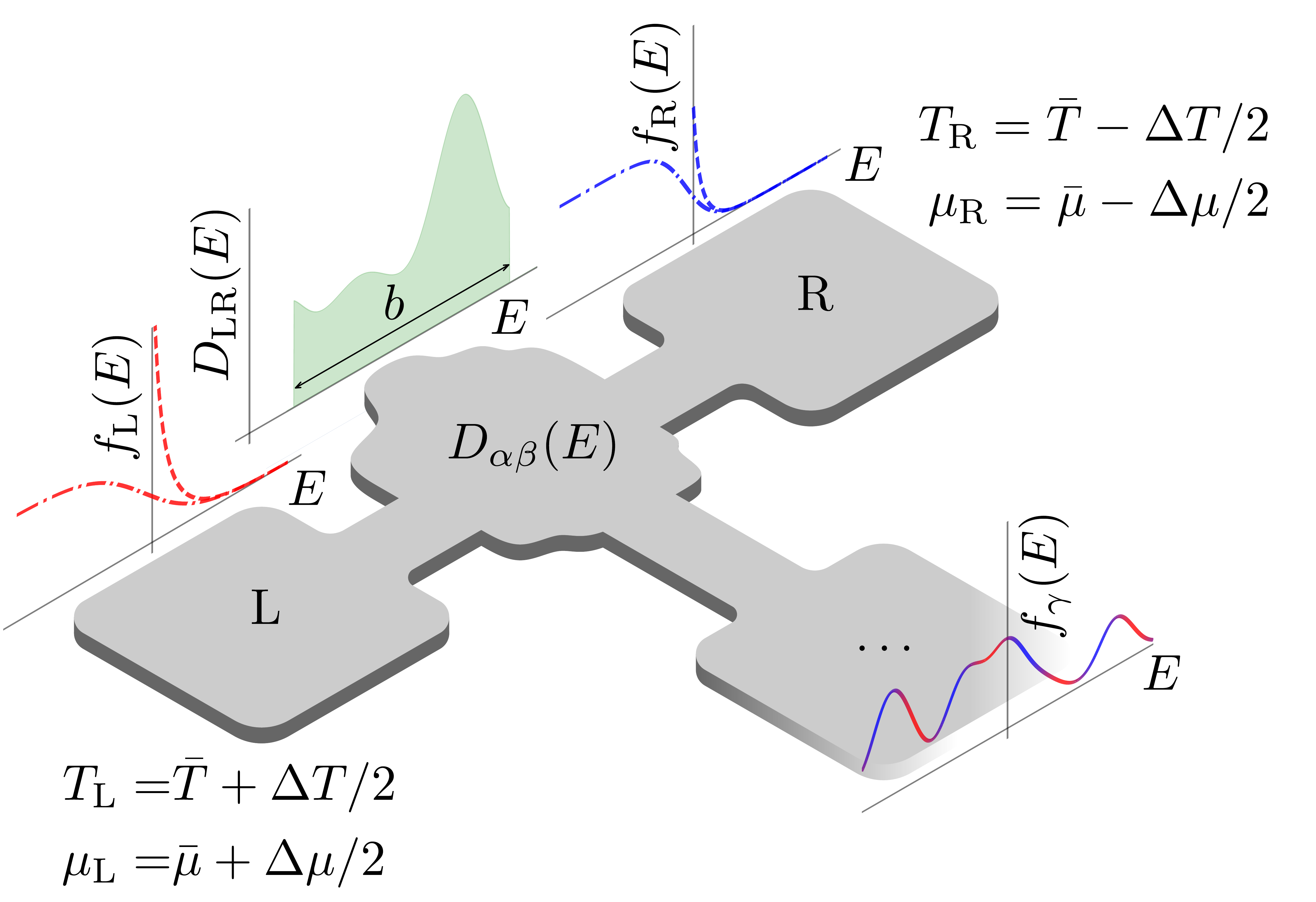}
    \caption{Sketch of a multi-terminal setup with contacts $\alpha=\text{L}, \text{R},...$, described by thermal fermionic (dashed-dotted lines in examples) or bosonic (dashed lines) distributions, or even nonthermal distributions ($f_\gamma$, full line). The central scattering region is characterized by transmission probabilities $D_{\alpha\beta}(E)$ (green area) with width $b$ in energy.}
    \label{fig:setup}
\end{figure}

This is the gap that we fill in this letter.
We analyze stationary quantum transport in a generic multi-terminal coherent setup, see Fig.~\ref{fig:setup}, where the contacts are described either by fermionic or bosonic, possibly nonthermal, distributions. For a large set of transport currents, such as particle, energy, or entropy currents that are also relevant in the context of thermodynamics, we show that the precision is limited by the particle-current fluctuations. In the weak-transmission limit---corresponding to negligible quantum correlations---the particle-current fluctuations can be interpreted as a \textit{local activity}. Thus, our findings represent a quantum-transport version of the KUR. 
However, the presence of quantum correlations impacts bosonic and fermionic systems strongly and very differently due to bunching or antibunching. For this general case where both classical and quantum fluctuations are present, we establish precision bounds, where the activity is replaced by a combination of measurable transport quantities.
We demonstrate our findings at the experimentally relevant examples of bosonic or fermionic two-terminal systems and show that the discovered bounds can indeed be tight far from equilibrium.


For our analysis of quantum transport in this generic multi-terminal setting, we resort to scattering theory~\cite{Blanter2000Sep,Moskalets2011Sep}, which is valid as long as the scattering region can effectively be described by a Hamiltonian that is quadratic in the field operators. This allows us to model currents and noise in any transport setting with interactions treated up to the mean-field level~\cite{Jauho1994Aug,Souza2008Oct,Haupt2010Oct}, including the possibility to model inelastic effects via B\"uttiker probes~\cite{Texier2000Sep}.
The setup shown in Fig.~\ref{fig:setup} is hence characterized by arbitrary transmission probabilities $D_{\alpha\beta}(E)$ between any contacts $\alpha,\beta$; no time-reversal symmetry is required.
Contacts are either described by standard Bose or Fermi functions, or even by nonthermal distributions, meaning that they need not be described by a temperature or an (electro-)chemical potential. Indeed, the only constraint on these distributions is that they fulfill $0\leq f_\alpha(E)$ for bosons and $0\leq f_\alpha(E)\leq 1$ for fermions, imposed by exchange statistics. 
With these ingredients, we write the currents of interest into contact $\alpha$ as
\begin{equation}\label{eq:current}
    I^{(\nu)}_\alpha = \frac{1}{h}\int_0^\infty\!\!\!\! dE x^{(\nu)}_\alpha\sum_\beta D_{\alpha\beta}(E)\left[f_\beta(E)-f_\alpha(E)\right].
\end{equation}
Here, the superscript $\nu$ indicates the type of current. The real-valued variable $x^{(\nu)}_\alpha$ for particle currents, $\nu=N$, is $x^{(N)}_\alpha=1$, for energy currents, $\nu=E$, it is $x^{(E)}_\alpha=E$, and for entropy currents~\cite{Deghi2020Jul,Acciai2024Feb}, $\nu=\Sigma$, it is $x^{(\Sigma)}_\alpha=k_\mathrm{B} \log \left[(1\pm f_\alpha(E))/f_\alpha(E)\right]$. From now on, the upper and lower signs refer to bosonic and fermionic systems, respectively. 
We are particularly interested in the precision of these currents. We therefore evaluate their fluctuations, focusing on the zero-frequency autocorrelator $S_{\alpha\alpha}^{(\nu)}=\int dt \langle \delta\hat{I}_\alpha^{(\nu)}(t)\delta\hat{I}_\alpha^{(\nu)}(0)\rangle$, with $\delta\hat{I}_\alpha^{(\nu)}(t)=\hat{I}_\alpha^{(\nu)}(t)-I_\alpha^{(\nu)}$ and $I_\alpha^{(\nu)}=\langle\hat{I}_\alpha^{(\nu)}(t)\rangle$. See Ref.~\onlinecite{Acciai2024Feb} for a detailed treatment of the entropy-current fluctuations. 
Most previous KUR studies deal with classical processes and for comparison it is therefore helpful to divide the full fluctuations into a ``classical" and a ``quantum" contribution, $S_{\alpha\alpha}^{(\nu)}=S_{\alpha\alpha}^{(\nu)\mathrm{cl}}+S_{\alpha\alpha}^{(\nu)\mathrm{qu}}$. The classical contribution
\begin{subequations}
\label{eq:S}
\begin{equation}\label{eq:Scl}
    S_{\alpha\alpha}^{(\nu)\mathrm{cl}}=\frac{1}{h}\int dE [x^{(\nu)}_{\alpha}(E)]^2 \biggr\{\sum_{\beta\neq\alpha}D_{\alpha\beta}( F^\pm_{\alpha \beta} + F^\pm_{\beta \alpha}) \biggr\} 
\end{equation}
is linear in the transmission probabilities, and we defined $F^\pm_{\alpha\beta}\equiv f_\alpha(1\pm f_\beta)$.
While quantum properties can enter {$D_{\alpha\beta}(E)$} via \textit{interference} effects, the classical part of the fluctuations only contains ``single-particle" effects due to the linear dependence on $D_{\alpha\beta}(E)$. Quantum \textit{correlations} are, in contrast, included in the quantum contribution
\begin{equation}\label{eq:Squ}
    S_{\alpha\alpha}^{(\nu)\mathrm{qu}}= \pm\frac{1}{h}\int dE [x^{(\nu)}_{\alpha}(E)]^2 \biggr\{\sum_{\beta\neq\alpha}D_{\alpha\beta} (f_\alpha-f_\beta) \biggr\}^2 ,
\end{equation}
\end{subequations}
which is quadratic in the transmission probabilities. The latter is purely due to nonequilibrium, i.e., it vanishes for $f_\alpha=f_\beta$. Importantly, it has opposite signs for bosonic and fermionic systems, meaning that it reduces the total noise for fermions while increasing it for bosons. We emphasize that while the splitting into classical and quantum fluctuations is useful from a mathematical standpoint, in experiments, the full fluctuations are measured.

\begin{figure*}[t]
    \centering
    \includegraphics[width=6.6in]{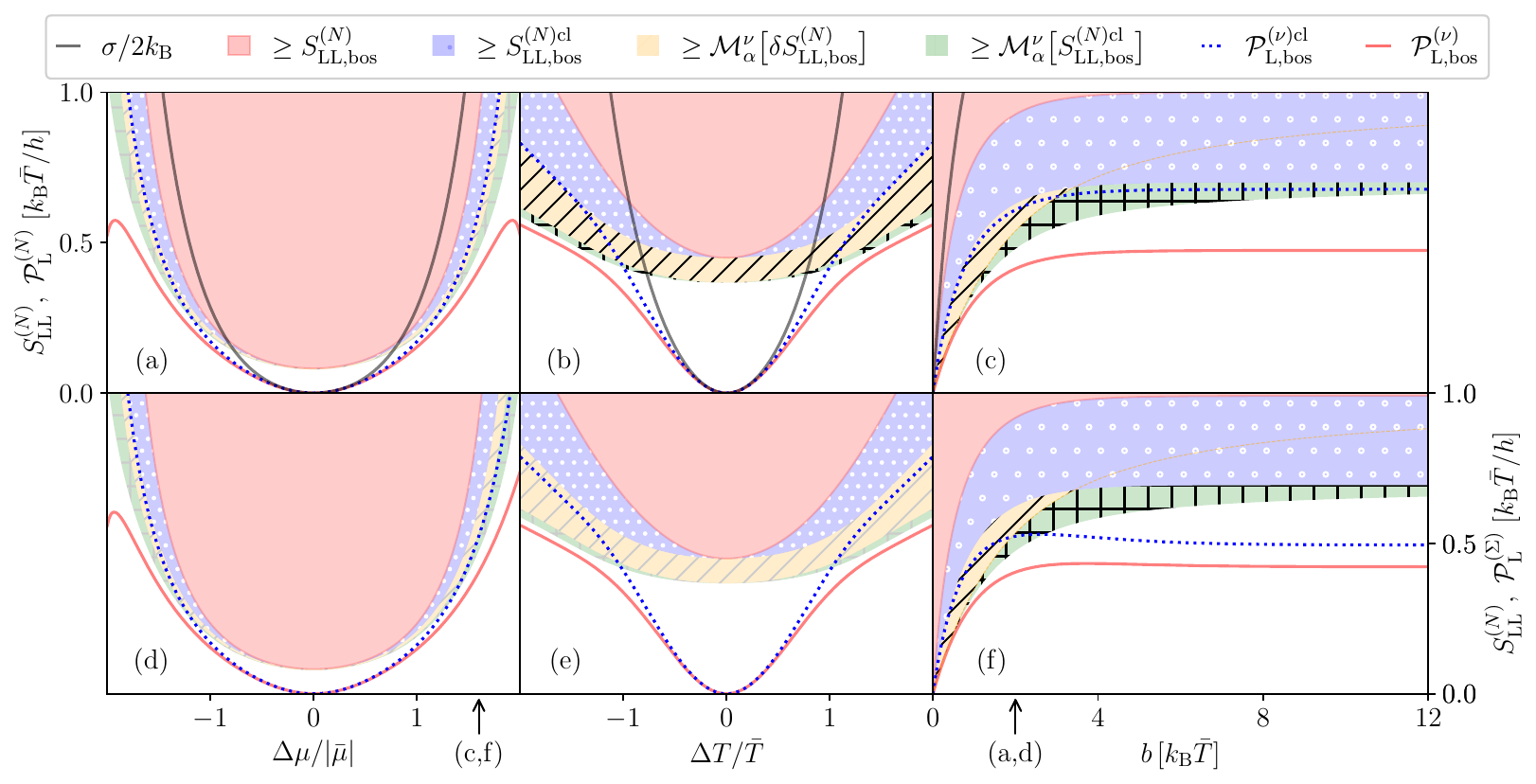}
    \caption{KUR and KURL, see bounds~\eqref{eq:cl-bound}, \eqref{eq:boson-full-weak}, \eqref{eq:half_boson_bound} and~\eqref{eq:full_boson_bound}, in a bosonic two-terminal system.
    Red and blue lines show precision functions, filled areas are the regions excluded by the bounds. Black lines show the total entropy production $\sigma=I^{(\Sigma)}_\text{L}+I^{(\Sigma)}_\text{R}$ constraining precision via the TUR. 
     Upper/lower row: bounds on \textit{particle}-/\textit{entropy}-current precision as function of (a,d) potential bias at $\Delta T=0$, (b,e) temperature bias at $\Delta \mu=0$, and (c,f) the bandwidth $b$.
    In panels (a,b,d,e), we choose $B_\mathrm{L}^{(\nu)}=b=2k_\text{B} \bar{T}$, panels (a,c,d,f) have $\bar{\mu}=-3k_\text{B}\bar{T}$ with $\Delta T=0$, panels (b,e) have $\bar{\mu}=-1.5k_\text{B}\bar{T}$, $\Delta \mu=0$, and (c,f) have $\Delta \mu=1.6|\bar{\mu}|$. Arrows show values of bandwidth and potential bias chosen for different plots. For all plots, $D_0=1$ and $E_0=0.1k_\text{B}\bar{T}$.}  
    \label{fig:boson_bound}
\end{figure*}

We start by analyzing the classical noise alone, which is the dominant contribution in the limit of weak transmissions $D_{\alpha\beta}(E)\ll1$ or for small biases, namely where $|f_\alpha-f_\beta|$ is small.
In order to establish bounds on this classical noise,
we use the inequalities
\begin{subequations}\label{eq:in}
    \begin{eqnarray}
    x^2+\frac{1}{4} & \geq &  |x|,\label{eq:inx}\\
    F^\pm_{\alpha\beta}(E)+F^\pm_{\beta\alpha}(E) & \geq & |f_\alpha(E)-f_\beta(E)|,\label{eq:inF}
\end{eqnarray}
\end{subequations}
which are valid for an arbitrary real number $x$ [(\ref{eq:inx})] and for bosonic as well as for fermionic arbitrary (possibly non-thermal) distribution $f_\alpha(E)$ and $F^\pm_{\alpha\beta}(E)$ [\eqref{eq:inF}].
Using inequalities~(\ref{eq:in}), we find~\cite{supp,Acciai2024Feb} the semidefinite positive quadratic form $S^{(\nu)\text{cl}}_{\alpha\alpha}+ \frac{y^2}{4}S^{(N)\text{cl}}_{\alpha\alpha} - y|I^{(\nu)}_\alpha|\geq0,
$ relating noise and current,
where $y$ has units of $x_\alpha^{(\nu)}$ and takes on positive, real values. 
The optimal value of $y$ that minimizes the quadratic form is $y=2 |I^{(\nu)}_\alpha|/S^{(N)\text{cl}}_{\alpha\alpha}$, which gives the bound on the classical precision $\P_\alpha^{(\nu)\text{cl}}$, 
\begin{equation}\label{eq:cl-bound}
    S^{(N)\mathrm{cl}}_{\alpha\alpha}\geq\frac{\left(I^{(\nu)}_\alpha\right)^2}{S^{(\nu)\mathrm{cl}}_{\alpha\alpha}}\equiv\P_\alpha^{(\nu)\text{cl}}. 
\end{equation}
It is valid for any current\footnote{For particle currents, this bound simply implies that classical noise is superpoissonian~\cite{Safi2014Jan}.}, $I^{(\nu)}_\alpha$, in either bosonic or fermionic, possibly nonthermal systems \cite{supp}, whenever the current and the noise can be written in the form of Eqs.~(\ref{eq:current}, \ref{eq:S}), respectively.
This bound, which can be interpreted as a \textit{kinetic uncertainty relation} is the first central result of this work.  
{For} this bound~\eqref{eq:cl-bound} on the \textit{classical} component of the noise, {we establish} a direct relation between the particle-current noise and the \textit{local activity} $\mathcal{K}_\alpha$ with respect to contact $\alpha$ of the system. See Ref.~\cite{Blasi2023Dec,Nishiyama2024Apr} for connections between fluctuations and activity in the weak transmission regime or for closed systems. 
We identify this local activity as
\begin{equation}\label{eq:activity}
S_{\alpha\alpha}^{(N)\mathrm{cl}}=\Gamma_{\alpha}^\to+\Gamma_{\alpha}^
\gets \equiv \mathcal{K}_\alpha\ ,
\end{equation}
with $\Gamma_{\alpha}^\to=\frac1h\sum_{\beta\neq\alpha}\int dE D_{\alpha\beta}(E)F_{\alpha\beta}^\pm(E)$ and $\Gamma_{\alpha}^\gets=\frac1h\sum_{\beta\neq\alpha}\int dE D_{\alpha\beta}(E)F_{\beta\alpha}^\pm(E)$.
The difference of these rates equals the particle current defined in Eq.~\eqref{eq:current}, and their sum in Eq.~\eqref{eq:activity} is a measure of how active the system is with respect to particle exchange with contact $\alpha$. The activity, $\mathcal{K}_\alpha$, hence limits the precision of a current as $\mathcal{K}_\alpha\geq \P_\alpha^{(\nu)\text{cl}}$.
The result for this KUR bound is shown {in blue} in Fig.~\ref{fig:boson_bound}, for a two-terminal conductor with thermal bosonic distribution functions with $T_\mathrm{L}=\bar{T}+\Delta T/2,T_\mathrm{R}=\bar{T}-\Delta T/2,\mu_\mathrm{L}=\bar{\mu}+\Delta \mu/2,\mu_\mathrm{R}=\bar{\mu}-\Delta \mu/2$, and with a boxcar-shaped transmission probability $D_{\mathrm{LR}}(E)\equiv D(E)=D_0\left[\theta(E-E_0)-\theta(E-E_0-b)\right]$.
Here, $b$ is the width of the transmission window and $E_0$ is the onset of the region with constant transmission $D_0\neq 0$.
Transmission functions approaching this boxcar shape can be realized, e.g., by wavelength-selective mirrors for optical systems, by multipole Purcell filters in circuit QED~\cite{Yan2023Sep}, or by chains of quantum dots~\cite{Whitney2015Mar} in electronic conductors.
We show results both for the particle-current precision $\P^{(N)\mathrm{cl}}_\mathrm{L}$ and the entropy-current precision $\P^{(\Sigma)\mathrm{cl}}_\mathrm{L}$ (dotted {blue} lines), as compared to the bound set by the activity $\mathcal{K}_\mathrm{L}$ (where the region excluded by $\mathcal{K}_\mathrm{L}$ is colored in light {blue}). 
Note that since we here choose the contact distributions to be thermal, entropy currents and their fluctuations are directly related to heat currents and their fluctuations via $x_\alpha^{\Sigma}\rightarrow (E-\mu_\alpha)/T_\alpha$. 
The plots show that the classical precision $\P_\mathrm{L}^{(\nu)\text{cl}}$ is large when either of the biases is large or when the bandwidth approaches the voltage bias, namely when current and activity are large.
While for large biases, the precision approaches the bound set by the activity, a large bandwidth increases the distance between the precision and its bound, as is particularly visible for the entropy current in the lower row of Fig.~\ref{fig:boson_bound}. 
The reason for this is that~\eqref{eq:inF} {approaches equality} when the distribution functions differ strongly, one of them being zero; this situation coincides with the activity approaching the magnitude of the particle-current, $\mathcal{K}_\alpha\rightarrow \Gamma_\alpha^{\rightleftharpoons}$. However, an increased bandwidth $b>\Delta\mu,T_\alpha$ necessarily involves energy-intervals in which the thermal contact distribution functions approach each other.
The tightness of the \textit{contact-selective} KUR of Eq.~(\ref{eq:cl-bound}) in various parameter regimes differs from the kinetic uncertainty relations for Markovian quantum jump processes\footnote{The reason is that activity here is \textit{local}, i.e. related to charge-transport rates in \textit{a single} reservoir. For a 2-terminal setup, e.g., one finds that Eq.~\eqref{eq:cl-bound} is tight when $|I_L| \rightarrow |\Gamma^{\rightleftharpoons}_L|$ and $S_L \rightarrow |\Gamma^{\rightleftharpoons}_L|$. Summing instead the activities of the two contacts, one has $S_L+S_R=2|\Gamma^{\rightleftharpoons}_L|\geq |\Gamma^{\rightleftharpoons}_L|=\mathcal{P}_L$.}, which were recently demonstrated not to be tight~\cite{Prech2024Jun} or only when the activity gets modified~\cite{Macieszczak2024Jul}.
Note that while the KUR is in general tight far away from equilibrium, constraints set by entropy production (TUR)~\cite{Barato2015Apr,Horowitz2020Jan,Palmqvist2024}, indicated by the black line, are most relevant close to equilibrium.

In general, it is the \textit{full} noise---including both the classical \textit{and} the quantum part---that is experimentally accessible and that influences the precision. Since the quantum contribution to the noise has an opposite overall sign for fermionic compared to bosonic systems, meaningful extensions of~\eqref{eq:cl-bound} are of different nature for these two cases. 
Indeed, for bosonic systems, where the quantum contribution to the noise is always positive, one finds that the classical kinetic uncertainty relation~\eqref{eq:cl-bound} continues to hold for the full noise
\begin{equation}\label{eq:boson-full-weak}
    S^{(N)}_{\alpha\alpha,\mathrm{bos}} \geq \mathcal{K}_\alpha \geq\frac{\left(I^{(\nu)}_{\alpha,\mathrm{bos}}\right)^2}{S^{(\nu)}_{\alpha\alpha,\mathrm{bos}}}\equiv\P_{\alpha,\mathrm{bos}}^{(\nu)}.
\end{equation}
The reason for the inequality to still hold is that quantum fluctuations in bosonic systems decrease the precision on the right hand side of the inequality while increasing the particle-current fluctuations on the left hand side. While relation~\eqref{eq:boson-full-weak} has the advantage of being valid for the full, experimentally accessible fluctuations, it has the drawback of being a \textit{loose} bound beyond the classical limit, see the red lines and surfaces in Fig.~\ref{fig:boson_bound}. 


In order to find an insightful KUR bound for bosonic {quantum} systems , we identify a general function of measurable transport quantities, which is representative for the activity in the quantum limit. As a first step, we estimate the magnitude of the quantum contribution to the noise identifying the important role of the transport bandwidth. 
We note that the quantum-noise integrand, Eq.~(\ref{eq:Squ}), equals the square of the current integrand, Eq.~\eqref{eq:current}. This integrand can be trivially extended by the indicator function of its support, namely $\zeta_\alpha^\nu (E)  = 1$ if $E\in\text{supp}\{x_\alpha^{(\nu)}\sum_\beta D_{\alpha\beta}\left(f_\alpha-f_\beta\right)\}$ and $\zeta_\alpha^\nu (E)  = 0$ otherwise.
Using Cauchy-Schwarz inequality for square-integrable functions~\cite{supp}, we find a bound on the \textit{quantum contribution} to the noise only,
\begin{equation}\label{eq:upper-qu-bound}
S_{\alpha\alpha,\mathrm{bos}}^{(\nu)\mathrm{qu}}\geq\frac{h}{B_\alpha^\nu}\left(I^{(\nu)}_\alpha\right)^2.
\end{equation}
Apart from the average current $I^{(\nu)}_\alpha$, it involves the bandwidth $B_\alpha^\nu =  \int_0^\infty dE\zeta_\alpha^\nu (E)$.
This shows that quantum bunching effects also constrain the full precision, $B_\alpha^\nu/h\geq(I^{(\nu)}_\alpha)^2/S_{\alpha\alpha,\mathrm{bos}}^{(\nu)\mathrm{qu}}\geq (I^{(\nu)}_\alpha)^2/S_{\alpha\alpha,\mathrm{bos}}^{(\nu)}$. 
The intuitive reason for this is that the smaller the bandwidth, the larger the relative weight of equal-energy states subject to bunching. Consequently, a small bandwidth leads to large quantum noise at a given current and hence to a reduced precision. Thus, to optimize precision in bosonic systems, it is beneficial to spread out transport over a large energy interval. Infinite bandwidth would make the bound on the quantum noise trivial; however, in a realistic setup, we always expect the energy interval contributing to transport to be finite.
Inequality~\eqref{eq:upper-qu-bound} reaches equality when the current integrand is constant over the bandwidth, e.g., when the bandwidth is much smaller than the contact temperatures. 

Combining \eqref{eq:cl-bound} and \eqref{eq:upper-qu-bound}, we capture the limit on the full precision, ${\mathcal{P}}_{\alpha,\mathrm{bos}}^{(\nu)}$, by a function of the classical particle-current noise, $\mathcal{K}_{\alpha,\mathrm{bos}}={S}^{(N)\text{cl}}_{\alpha\alpha,\mathrm{bos}}$, and of the bandwidth,
\begin{eqnarray}
     {\mathcal{M}_\alpha^\nu \left[\mathcal{K}_{\alpha,\mathrm{bos}} \right]\equiv \frac{\mathcal{K}_{\alpha,\mathrm{bos}} }{1+\frac{h}{B^\nu_\alpha} \mathcal{K}_{\alpha,\mathrm{bos}} }} \geq {\mathcal{P}}_{\alpha,\mathrm{bos}}^{(\nu)}.\label{eq:half_boson_bound}
\end{eqnarray}
In this KUR-type bound,  the structure of the function $\mathcal{M}^\nu_\alpha[x]$ takes into account the reduction of precision due to bunching in the presence of a limited bandwidth, while the classical particle-current noise still serves as an activity quantifying the number of single-particle transfers in and out of reservoir $\alpha$.
The bound~\eqref{eq:half_boson_bound} on the full precision is displayed as the shaded green areas in Fig.~\ref{fig:boson_bound}, where it shows clear improvements with respect to the loose bound~(\ref{eq:boson-full-weak}). In particular, it is tight for small bandwidths.

The drawback of the KUR bound~\eqref{eq:half_boson_bound} is that the classical activity is not experimentally accessible since it comprises only the classical noise contribution. As a next step, we therefore estimate the activity from measurable observables~\cite{Kobayashi2021Sep} (current, full particle-current noise, and bandwidth) exploiting the relation~\eqref{eq:upper-qu-bound}. Concretely, we introduce the function $\delta S^{(\nu)}_{\alpha \alpha,\mathrm{bos}} \equiv S_{\alpha\alpha,\mathrm{bos}}^{(\nu)}-h(I^{(\nu)}_\alpha)^2/B_\alpha^\nu$, which we find to be a general upper bound to the activity and which equals the classical activity when the bandwidth is small compared to temperature, even far from equilibrium.
We find the following KUR-like bound (KURL)
\begin{eqnarray}
     { \mathcal{M}_\alpha^\nu \left[\delta {S}^{(N)}_{\alpha\alpha,\mathrm{bos}} \right] \geq\mathcal{M}_\alpha^\nu \left[\mathcal{K}_{\alpha,\mathrm{bos}} \right] \geq {\mathcal{P}}_{\alpha, \mathrm{bos}}^{(\nu)},} \label{eq:full_boson_bound}
\end{eqnarray}
 allowing us to limit the precision of an arbitrary current $I_\alpha^{(\nu)}$ in terms of measurable particle-transport observables, which give an estimate of the activity in the full quantum case. Together with the KUR  bound~\eqref{eq:half_boson_bound} in terms of classical activity it constitutes the second main result of this paper. The KURL bound~\eqref{eq:full_boson_bound} is shown in Fig.~\ref{fig:boson_bound} as shaded yellow areas. Notably, there is a crossing in panels (c,f) between the activity setting the bound in~\eqref{eq:cl-bound} and the function $ \mathcal{M}_\alpha^\nu \left[\delta {S}^{(N)}_{\alpha\alpha,\mathrm{bos}} \right]$ setting the bound in~\eqref{eq:full_boson_bound}. This is because $\mathcal{M}_\alpha^\nu[x]$ accounts for bunching effects, relevant for small bandwidths, which the classical activity fails to do.
For small biases $\Delta T$ or $\Delta \mu$, the quantum part of the noise is negligible, meaning that the classical and full precisions, $\P_\alpha^{(\nu)\text{cl}}$ and $\P_\alpha^{(\nu)}$, are the same and ${S}^{(\nu)\mathrm{cl}}_{\alpha\alpha,\text{bos}}$ and the respective bounds coincide.

For \textit{fermionic} systems, an inequality of the type of~\eqref{eq:boson-full-weak} is generally not valid since it can be broken by the quantum contribution to the noise.
Indeed, anti-bunching decreases the particle-current noise entering~\eqref{eq:cl-bound}, while increasing precision. Bounds accounting for anti-bunching in fermionic systems, expressed in terms of experimentally accessible quantities are hence desirable. 
In order to establish a relation between the fermionic quantum noise and the current, we estimate~\cite{Acciai2024Feb},
\begin{equation}
    S^{(\nu)\text{qu}}_{\alpha \alpha,\mathrm{fer}} \geq -\frac{1}{h} \int dE [x^{(\nu)}_{\alpha}]^2[1-D_{\alpha\alpha}]\sum_{\beta\neq\alpha}D_{\alpha \beta} |F^-_{\beta\alpha}+F^-_{\alpha\beta}| 
\end{equation}
which approaches equality far from equilibrium, namely when the fermionic distribution functions differ maximally from each other. Introducing the minimum reflection probability inside the bias window, namely in the energy interval $A$ where transport happens, $R_\alpha \equiv \inf_{E\in A}D_{\alpha\alpha}(E)$, we find a fermionic KURL for the full fluctuations~\cite{supp}
\begin{equation}\label{eq:Full bound fermion}
{\frac{1}{R_\alpha}\frac{S^{(N)}_{\alpha\alpha,\mathrm{fer}}}{R_\alpha}\geq \frac{1}{R_\alpha} S^{(N)\mathrm{cl}}_{\alpha\alpha,\mathrm{fer}}}\geq\frac{\left(I^{(\nu)}_\alpha\right)^2}{S^{(\nu)}_{\alpha\alpha,\mathrm{fer}}} \equiv \mathcal{P}^{(\nu)}_{\alpha,\mathrm{fer}}.
\end{equation}
In the limit of vanishing reflection probabilities, the precision is \textit{not} bounded. Indeed, for a fully transmitting scattering region and large potential bias, currents are known to be noiseless due to fermionic antibunching, leading e.g., to a breaking of the TUR~\cite{Brandner2018Mar,Agarwalla2018Oct,Timpanaro2023Mar,Tesser2024May}. Similarly to the bosonic case, $ S^{(N)\mathrm{cl}}_{\alpha\alpha,\mathrm{fer}}$ still takes the role of an activity quantifying the number of single-particle transfers in and out of reservoir $\alpha$. The upper bound $S^{(N)\mathrm{cl}}_{\alpha\alpha,\mathrm{fer}} /R_\alpha$ is thus a function of the classical activity and a direct extension of the KUR accounting for antibunching. Again, we aim to express this bound in terms of the measurable, full particle-current noise, and we therefore exploit $S^{(N)}_{\alpha\alpha,\mathrm{fer}} /R_\alpha\geq S^{(N)\mathrm{cl}}_{\alpha\alpha,\mathrm{fer}} $. The leftmost bound of~\eqref{eq:Full bound fermion} is hence both taking into account the effect of antibunching as well as an estimate of the true activity from the full fluctuations. 

\begin{figure}[t!]
    \centering
\includegraphics[width=3.3in]{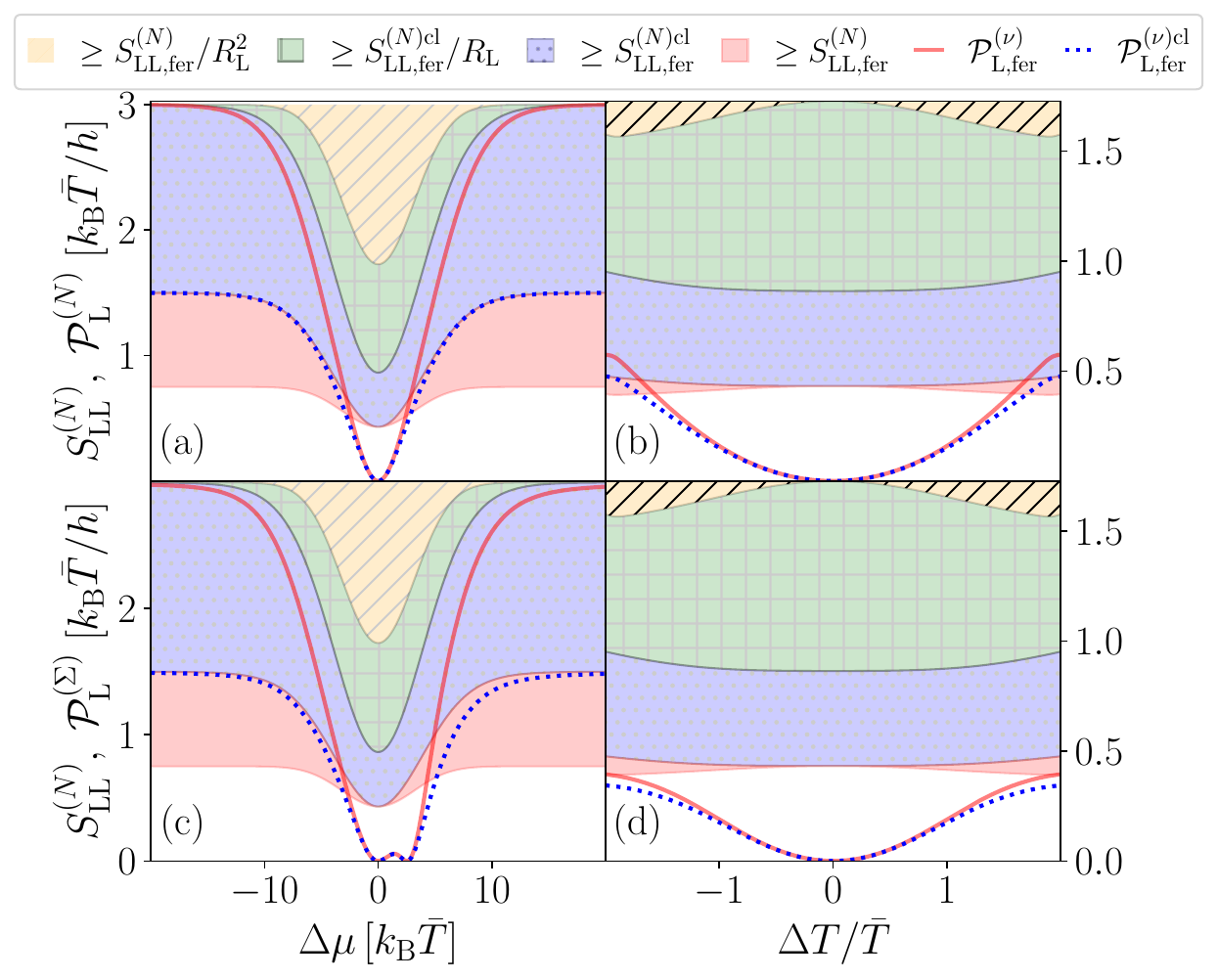}
    \caption{KUR in a fermionic two-terminal system.
    The lines show the precision functions and the filled areas the regions excluded by the bounds, see bounds~(\ref{eq:cl-bound},\ref{eq:Full bound fermion}).
    Upper/lower row: bounds on \textit{particle}-/\textit{entropy}-current precision as function of (a,c) chemical potential bias $\Delta\mu$ and (b,d) temperature bias $\Delta T$. In all panels we use $\bar{\mu}\equiv0$ and $D_0=0.5$, $E_0=0.1k_\mathrm{B}\bar{T}$, $b=1k_\mathrm{B}\bar{T}$ for the boxcar transmission function. Panels (a,c) have $\Delta T=0$, while panels (b,d) have $\Delta \mu=0$.}
    \label{fig:fermionic_bound}
\end{figure}

The fermionic KUR for classical fluctuations,~\eqref{eq:cl-bound}, as well as the full fermionic KURL,~\eqref{eq:Full bound fermion}, are shown in Fig.~\ref{fig:fermionic_bound} for particle- and entropy-current precision in a thermal two-terminal system with boxcar-shaped transmission probability. 
The blue surface bounds the classical precision from above, see~\eqref{eq:cl-bound}. 
However, in contrast to the bosonic case, the full particle-current noise is decreased with respect to the classical one. 
As soon as temperature or potential bias are of the order of the average temperature, the precision exceeds the classical, or even the full, particle-current noise, breaking the classical KUR as expected. The precision is instead bounded by the particle-current noise modified by the minimum reflection probability $R_\mathrm{L}$, as given in~\eqref{eq:Full bound fermion}. These bounds become tight for large potential biases, when one of the distributions is zero inside the interval of nonzero transmission. In addition, a transmission function which is flat within the relevant energy window is required to make~\eqref{eq:Full bound fermion} tight. The main difference in the precision for the entropy current (lower row) compared to the one for the particle current (upper row) is the occurrence of an additional zero. It appears at finite bias, when $\mu_\mathrm{L}$ is aligned with the center of the boxcar transmission such that the full entropy production happens in the right contact.

In summary, we have established general precision bounds for quantum transport valid for any setup that can effectively be described by scattering theory. While these bounds take the form of a kinetic uncertainty relation bounded by a \textit{local} activity in the classical limit, we here provide analogous constraints for the precision in the presence of quantum fluctuations. These bounds are given by functions of the classical activity accounting for quantum statistics at arbitrary transmission, thereby filling a gap that can not be addressed by standardly used weak-coupling approaches. As a result of quantum statistics, the precision of bosonic systems benefits from large bandwidths, and the precision of fermionic systems benefits from low reflections. We further reformulate these quantum KURs in terms of measurable particle-current noise and transmission properties. The obtained bounds are expected to serve as guidelines for the design of any multi-terminal device aiming at precision. Additionally, the bounds could serve as inference tools for estimating fluctuations in heat and entropy currents.

\textit{Acknowledgments.---}
We thank Gabriel Landi for helpful discussions. We gratefully acknowledge funding from the Knut and Alice Wallenberg foundation via the fellowship program (L.T. and J.S.) and the European Research Council (ERC) under the European Union’s Horizon Europe research and innovation program (101088169/NanoRecycle) (D.P. and J.S.).

\bibliography{refs.bib}
\end{document}


\title{Kinetic uncertainty relations for quantum transport:\\ Supplementary material }

\author{Didrik Palmqvist}
	\affiliation{Department of Microtechnology and Nanoscience (MC2), Chalmers University of Technology, S-412 96 G\"oteborg, Sweden}

\author{Ludovico Tesser}
	\affiliation{Department of Microtechnology and Nanoscience (MC2), Chalmers University of Technology, S-412 96 G\"oteborg, Sweden}

	\author{Janine Splettstoesser}
	\affiliation{Department of Microtechnology and Nanoscience (MC2), Chalmers University of Technology, S-412 96 G\"oteborg, Sweden}
	
\date{\today}
\maketitle

\makeatletter\@input{xx.tex}\makeatother

\section{Classical noise bounds}\label{sec:cl noise bounds}
To show that the KUR holds for the classical part of the fluctuations, we begin by proving the inequalities~\eqref{eq:inx} and~\eqref{eq:inF} from the main paper.  Inequality~\eqref{eq:inx} follows from the quadratic form 
\begin{equation}\label{eq:supp inx}
    x^2 -|x|+\frac14 = \left(|x|-\frac12\right)^2\geq 0\quad\Rightarrow\quad x^2+\frac14\geq |x|.
\end{equation}
Inequality~\eqref{eq:inF} stems from 
\begin{equation} \label{eq:supp inF}
|x+y| \geq |x-y|,\quad \text{for}\quad x,y\geq0    
\end{equation}
with $x = F_{\alpha\beta}^\pm =f_\alpha(1\pm f_\beta)$ and $y=F_{\beta\alpha}^\pm =f_\beta(1\pm f_\alpha)$ for bosons ($+$) and fermions ($-$). Note that the conditions $f_\alpha\geq 0$ for bosons and $0\leq f_\alpha\leq 1$ for fermions guarantee that $F_{\alpha\beta}^\pm\geq 0$.
Using now the inequalities~\eqref{eq:inx} and \eqref{eq:inF} in the integrand of the classical component of the noise in Eq.~\eqref{eq:Scl} in the main paper, we find 
\begin{align}
    \frac{1}{h} \bigg(\left[\frac{x^{\nu}_\alpha(E)}{y}\right]^2 +\frac{1}{4}\bigg) \biggr\{\sum_{\beta\neq\alpha} D_{\alpha\beta}( F^\pm_{\alpha \beta} + F^\pm_{\beta \alpha}) \biggr\} \geq \frac{1}{h} \left|\frac{x^{\nu}_\alpha(E)}{y}\right| \biggr\{\sum_{\beta\neq\alpha} D_{\alpha\beta}|f_\beta
    -f_\alpha| \biggr\},
\label{eq:Step1 cl KUR}
\end{align}
where we introduced a parameter $y\geq 0$ such that $x^{\nu}_\alpha(E)/y$ is dimensionless.
Since this holds at any energy $E$, by integration we find
\begin{equation}\label{eq:Step2 cl KUR}
  \frac{1}{y^2}S^{(\nu)\text{cl}}_{\alpha\alpha}+ \frac{1}{4}S^{(N)\text{cl}}_{\alpha\alpha}\geq \frac1h\int dE   \left|\frac{x^{\nu}_\alpha(E)}{y}\right| \biggr\{\sum_{\beta\neq\alpha} D_{\alpha\beta}|f_\beta
    -f_\alpha| \biggr\} \geq \frac{1}{y}\left|I^{(\nu)}_\alpha\right|,
\end{equation}
where we moved the absolute value outside the integral in the last inequality. This yields a positive quadratic form
\begin{equation}\label{eq:Step3 cl KUR}
  S^{(\nu)\text{cl}}_{\alpha\alpha}+ \frac{y^2}{4}S^{(N)\text{cl}}_{\alpha\alpha}- y\left|I^{(\nu)}_\alpha\right|\geq0.
\end{equation}
The bound~\eqref{eq:Step3 cl KUR} holds for any $y$. We are interested in finding a constraint  between fluctuations and currents that is as tight as possible, so we minimize the form. Taking a derivative with respect to $y$ and setting it to zero, we find
\begin{equation}\label{eq:Step4 cl KUR}
   \frac{y}{2}S^{(N)\text{cl}}_{\alpha\alpha}- \left|I^{(\nu)}_\alpha\right|=0 \implies y = 2\frac{\left|I^{(\nu)}_\alpha\right|}{S^{(N)\text{cl}}_{\alpha\alpha}}.
\end{equation}
Inserting the optimal value of $y$ yields 
\begin{equation}\label{eq:Step6 cl KUR}
S^{(N)\mathrm{cl}}_{\alpha\alpha}\geq\frac{\left(I^{(\nu)}_\alpha\right)^2}{S^{(\nu)\mathrm{cl}}_{\alpha\alpha}} .
\end{equation}
Thus, we have proven that the classical part of the particle-current fluctuations limits the precision in \textit{any} current with respect to its classical fluctuations.
We interpret this as a kinetic uncertainty relation applied on the classical contribution to the noise, based on the connection between activity and classical noise given by Eq.~\eqref{eq:activity} in the main text. 

\section{Quantum noise bounds}

\subsection{Noise bounds for bosonic systems}

In bosonic systems, the classical KUR~\eqref{eq:cl-bound} can be extended to the full measurable fluctuations
\begin{equation}
S^{(N)}_{\alpha\alpha,\text{bos}}\geq S^{(N)\mathrm{cl}}_{\alpha\alpha,\text{bos}}\geq\frac{\left(I^{(\nu)}_{\alpha,\text{bos}}\right)^2}{S^{(\nu)\mathrm{cl}}_{\alpha\alpha,\text{bos}}} \geq \frac{\left(I^{(\nu)}_{\alpha,\text{bos}}\right)^2}{S^{(\nu)}_{\alpha\alpha,\text{bos}}}.
\end{equation}
However, this bound is generally not tight, particularly far from equilibrium, where the quantum part of the fluctuations becomes the dominant contribution to the overall noise. To find a stricter constraint on the precision, we develop additional relations treating the quantum part of the fluctuations. This yields at the same time insights into the differences between precision in fermionic and bosonic systems since exchange statistics affect two-particle processes.
We define
\begin{equation}
    g_\alpha(E) \equiv \sum_{\beta\neq \alpha} D_{\alpha\beta} (f_\beta- f_\alpha),\label{eq:g}
\end{equation}
and use the indicator function $\zeta^{(\nu)}_\alpha(E)$ as defined before~\eqref{eq:upper-qu-bound} in the main text. This allows us to express
\begin{align}
    \big|{I^{(\nu)}_{\alpha}}\big| &\leq \frac{1}{h}\int_0 ^{\infty} dE| x^{\nu}_{\alpha}(E) g_\alpha(E) \zeta^{(\nu)}_\alpha(E)|, \label{eq: CSI c-weighted current} \\
    S^{(\nu)\text{qu}}_{\alpha \alpha,\text{bos}} &= \frac{2}{h}\int_0 ^{\infty} dE \left( x^{\nu}_{\alpha}(E) g_\alpha (E)\right)^2.
\end{align}
By applying the Cauchy-Schwarz integral inequality, we find
\begin{align}\label{eq:CS deriv}
    {h^2}  \big|{I^{(\nu)}_{\alpha,\text{bos}}}\big|^2 &\leq \left(\int_0 ^{\infty} dE {|x^{\nu}_{\alpha}(E)} g_\alpha(E) \zeta^{(\nu)}_\alpha(E)|\right)^2 \nonumber
    \\ &\leq \int_0 ^{\infty} dE \left({x^{\nu}_{\alpha}(E)} g_\alpha(E) \right)^2 \int_0 ^{\infty} dE \left( \zeta^{(\nu)}_\alpha(E)\right)^2
    = {h}S^{(\nu) \text{qu}}_{\alpha \alpha,\text{bos}} \int_0 ^{\infty} dE  \zeta_\alpha^{(\nu)}(E) = h B^{\nu}_\alpha S^{(\nu) \text{qu}}_{\alpha \alpha,\text{bos}}.
\end{align}
Here, the bandwidth of transport $B^{\nu}_\alpha$ is introduced in the last step, as also defined below~(\ref{eq:upper-qu-bound}) in the main text. Rewriting this inequality, we find that the bandwidth constrains the precision in any current
\begin{equation}
    \frac{B^{\nu}_\alpha}{ h} \geq \frac{\big|{I^{(\nu)}_{\alpha,\text{bos}}}\big|^2 }{ S^{(\nu)\text{qu}}_{\alpha \alpha,\text{bos}}} \geq \frac{\big|{I^{(\nu)}_{\alpha,\text{bos}}}\big|^2 }{ S^{(\nu)}_{\alpha \alpha,\text{bos}}}. \label{eq: quantum precision bound}
\end{equation}
This inequality approaches equality whenever the integrand of the current,  $x^{\nu}_{\alpha}(E) g_\alpha(E)$, is constant. Physically, this indicates that to maximize the precision of a current with an energy-independent $x^{\nu}_{\alpha}$ for a given bandwidth, one should aim to distribute the particles evenly across the available energy spectrum, thereby avoiding bunching. 
For the energy current, one should instead aim at distributing the transmission of particles such that there are more low-energy particles, since bunching of high energy particles leads to greater fluctuations in the transmitted energy. This could be achieved by, e.g., non-thermal occupations $f_\alpha(E) = f_\alpha /E$.

Next, we combine~\eqref{eq:CS deriv} with the classical KUR for the particle current, yielding the compact result
\begin{equation}
    S^{(N)}_{\alpha \alpha,\text{bos}} \geq \left(1+ \frac{h}{B^{\nu}_\alpha}\big|{I^{(N)}_{\alpha,\text{bos}}}\big| \right) \big|{I^{(N)}_{\alpha,\text{bos}}}\big|.
\end{equation}
Substituting~\eqref{eq: quantum precision bound} into the KUR results in the KURL in
\begin{equation}\label{eq:KURL_bos}
    \left(S_{\alpha \alpha,\text{bos}}^{(N)}-\frac{ h}{B^{\nu}_{\alpha}} \big|{I^{(N)}_{\alpha,\text{bos}}}\big|^2 \right) \geq  \frac{ \big|{I^{(\nu)}_{\alpha,\text{bos}}}\big|^2}{\left(S_{\alpha \alpha,\text{bos}}^{(\nu)}-\frac{ h}{B^{\nu}_\alpha} \big|{I^{(\nu)}_{\alpha,\text{bos}}}\big|^2 \right)}.
\end{equation}
Re-expressing this bounds in terms of $\delta S^{(\nu)}_{\alpha \alpha}$ defined in the main text, we summarize the bosonic KUR-like bounds as follows,
\begin{equation}\label{eq: all inclusive}
S^{(N)}_{\alpha\alpha,\text{bos}}\geq \delta{S}^{(N)}_{\alpha\alpha,\text{bos}}\geq S^{(N)\mathrm{cl}}_{\alpha\alpha,\text{bos}}\geq\frac{\left(I^{(\nu)}_{\alpha,\text{bos}}\right)^2}{S^{(\nu)\mathrm{cl}}_{\alpha\alpha,\text{bos}}} \geq \frac{\left(I^{(\nu)}_{\alpha,\text{bos}}\right)^2}{\delta{S}^{(\nu)}_{\alpha\alpha,\text{bos}}} \geq \frac{\left(I^{(\nu)}_{\alpha,\text{bos}}\right)^2}{S^{(\nu)}_{\alpha\alpha,\text{bos}}}.
\end{equation} 
By using $\delta{S}^{(\nu)}_{\alpha\alpha,\text{bos}}$, we are hence able to find a stricter bound compared to utilizing the total fluctuations ${S}^{(\nu)}_{\alpha\alpha,\text{bos}}$. From~\eqref{eq: all inclusive} it is evident that it would be preferable to know the activity $\mathcal{K}_\alpha=S^{(N)\mathrm{cl}}_{\alpha\alpha}$ in order to constrain precision. The activity due to single particle transfers between reservoirs is, however, a difficult quantity to access experimentally in general. Instead, we utilize $\delta{S}^{(N)}_{\alpha\alpha}$ as an upper bound on the activity, which allows us to estimate it from experimentally accessible parameters, namely
\begin{equation}
    \delta{S}^{(N)}_{\alpha\alpha,\text{bos}} ={S}^{(N)}_{\alpha\alpha,\text{bos}}-\frac{ h}{B^{\nu}_\alpha} \big|{I^{(N)}_{\alpha,\text{bos}}}\big|^2\geq S^{(N)\mathrm{cl}}_{\alpha\alpha,\text{bos}},
\end{equation}
where the average currents, their fluctuations, and the bandwidth of transport appear.
For the full precision, we can also rewrite the bound~\eqref{eq:KURL_bos}, recovering the KURL~\eqref{eq:full_boson_bound} of the main text
 \begin{equation}
   \mathcal{P}_{\alpha,\text{bos}}^{(\nu)}=\frac{\left(I^{(\nu)}_{\alpha,\text{bos}}\right)^2}{S^{(\nu)}_{\alpha\alpha,\text{bos}}}\leq \frac{\delta{S}^{(N)}_{\alpha\alpha,\text{bos}}}{1+\frac{\delta{S}^{(N)}_{\alpha\alpha,\text{bos}}}{B^{\nu}_\alpha/h}} \equiv \mathcal{M}^\nu_\alpha \left[\delta{S}^{(N)}_{\alpha\alpha,\text{bos}}\right] .
 \end{equation}
\subsection{Noise bounds for fermionic systems}
In contrast to the bosonic case, when considering the full fluctuations in the fermionic case, the KUR breaks down
\begin{equation}
S^{(N)}_{\alpha\alpha,\text{fer}} \not\geq S^{(N)\mathrm{cl}}_{\alpha\alpha,\text{fer}}\not\geq \frac{\left(I^{(\nu)}_{\alpha,\text{fer}}\right)^2}{S^{(\nu)}_{\alpha\alpha,\text{fer}}}.
\end{equation}
Indeed, the negative sign of the fermionic quantum noise decreases the total fluctuations. This allows for higher precision, both when comparing to bosonic systems and to fermionic systems in the classical limit. It is, however, possible to derive less strict bounds that include the full measurable fluctuations.

To derive the bound~\eqref{eq:Full bound fermion} in the main text, we start from the inequality for the quantum contribution to the fluctuations~\cite{Acciai2024Feb},
\begin{equation}\label{app:eq:fermionic-qu-noise}
    S^{(\nu)\text{qu}}_{\alpha \alpha,\text{fer}} \geq -\frac{1}{h} \int dE [x^{\nu}_{\alpha}]^2[1-D_{\alpha\alpha}]\sum_{\beta\neq\alpha}D_{\alpha \beta} |f_\beta-f_\alpha|, 
\end{equation}
which stems from Jensen's inequality  
\begin{equation}
\left[\sum_{\beta\neq\alpha}\frac{D_{\alpha\beta}}{1-D_{\alpha\alpha}}(f_\alpha-f_\beta)\right]^2 \leq \sum_{\beta\neq\alpha}\frac{D_{\alpha\beta}}{1-D_{\alpha\alpha}}(f_\alpha-f_\beta)^2 \leq \sum_{\beta\neq\alpha}\frac{D_{\alpha\beta}}{1-D_{\alpha\alpha}}|f_\alpha-f_\beta|.
\end{equation}
Note that dividing by the factor $1-D_{\alpha\alpha}=\sum_{\beta\neq\alpha}D_{\alpha\beta}$ guarantees the normalization of the probabilities used in Jensen's inequality. In the last inequality we used that $|f_\alpha-f_\beta|\in[0,1]$.
Combining this with~\eqref{eq:supp inF} (equivalently~\eqref{eq:inF} from the main text) allows us to bound the full noise by the classical noise
\begin{equation}\label{eq:supp R bound}
   S^{(\nu) \mathrm{qu}}_{\alpha \alpha,\mathrm{fer}}\geq -(1-R_\alpha)S^{(\nu) \mathrm{cl}}_{\alpha \alpha,\mathrm{fer}}  \implies \frac{S^{(\nu)}_{\alpha \alpha,\mathrm{fer}}}{R_\alpha} \geq S^{(\nu) \mathrm{cl}}_{\alpha \alpha,\mathrm{fer}}.
\end{equation}
Here, we defined the minimum reflection probability inside the support $A$ of the current integrand 
\begin{equation}
    R_\alpha = \inf_{E\in A} D_{\alpha\alpha}(E).
\end{equation}
We are able to use~\eqref{eq:supp R bound} both in order to bound the full precision and to estimate the activity from the noise $S^{(N)}_{\alpha \alpha, \mathrm{fer}} /R_\alpha \geq S^{(N)\mathrm{cl}}_{\alpha \alpha, \mathrm{fer}} = \mathcal{K}_{\alpha,\mathrm{fer}}$.
When inserting~\eqref{eq:supp R bound} into~\eqref{eq:cl-bound}, we find a fermionic constraint on precision similar to the KUR
\begin{equation}\label{app:eq:fermionic-KURL}
\frac{1}{R_\alpha}\frac{S^{(N)}_{\alpha\alpha,\text{fer}}}{R_\alpha} \geq\frac{1}{{R_\alpha}}{S^{(N)\mathrm{cl}}_{\alpha\alpha,\text{fer}}}\geq\frac{\left(I^{(\nu)}_{\alpha,\text{fer}}\right)^2}{S^{(\nu)}_{\alpha\alpha,\text{fer}}} .
\end{equation}
This is the result of~\eqref{eq:Full bound fermion} in the main text. Note that as $R_\alpha \rightarrow 0$, this bound no longer poses a constraint on the precision in fermionic systems. Indeed, in the absence of partitioning, fluctuations can be reduced arbitrarily via anti-bunching.

\section{Additional examples of KUR}

\subsection{Constant average occupations}
\begin{figure}[h!]
 \centering \includegraphics[width=0.7\linewidth]{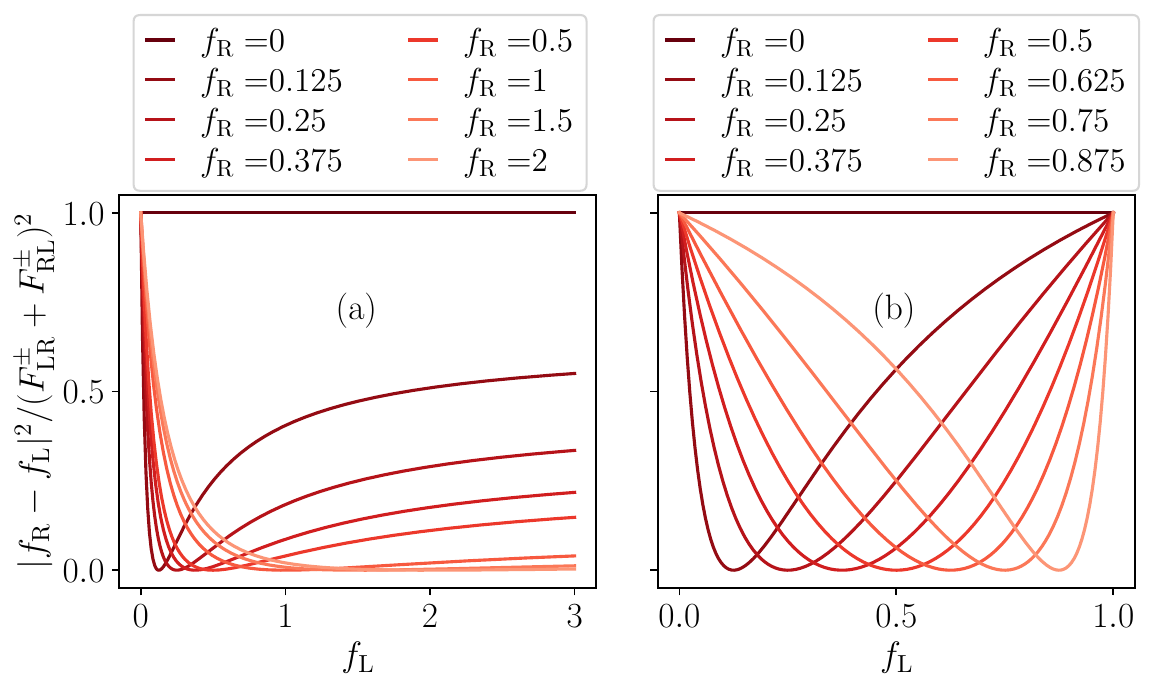}
 \captionsetup{justification=raggedright,singlelinecheck=false, font=small, width=0.9\textwidth} 
    \caption{Saturation of kinetic uncertainty relation-like bounds for two terminal system with constant average occupations $f_\text{L},f_\text{R}$ and a boxcar transmission. The lines show the ratio of Eq.~\eqref{eq: ratio constant f} as a function of $f_\text{L}$ for different values of $f_\text{R}$. Panel~(a) displays the bosonic bound and panel~(b) displays the fermionic case. Note that these results are independent of the parameters of the boxcar transmission.}
    \label{fig:constant f}
\end{figure}

As discussed in the main text, the bandwidth of transport plays an important role in the tightness of our bounds, which is in part due to on which scale the average occupations vary compared to the energy scale set by the bandwidth.
In fact, the bounds can be saturated in systems with constant average occupations for the precision of particle and entropy currents. 
To see this, we consider a two-terminal system with occupations $f_\text{L}$ and $f_\text{R}$ constant in energy and a boxcar transmission, for which bounds~(\ref{eq:cl-bound},\ref{eq:full_boson_bound}) presented in the main text reduce to~\eqref{eq:inF}. From this, it follows that the ratio between precision and activity simplifies to
\begin{equation}\label{eq: ratio constant f}
    1\geq\frac{\mathcal{P}^{(N,\Sigma)\text{cl}}_{\mathrm{L}}}{S_{\mathrm{L}\mathrm{L}}^{(N)\text{cl}}}=    \frac{{\mathcal{P}}^{(N,\Sigma)}_{\mathrm{L}}}{\mathcal{M}^{(N,\Sigma)}_{\mathrm{L}}\left[\delta{S}_{\mathrm{L}\mathrm{L}}^{(N)}\right]} = \frac{|f_\text{R}-f_\text{L}|^2}{(F_\text{RL}^\pm+F_\text{LR}^\pm)^2},
\end{equation}
independent of the parameters used for the boxcar transmission. For bosonic systems, this bound is tight whenever one of the reservoirs is empty, $f_\alpha=0$ while for fermions it is tight when one of the occupations fulfills $f_\alpha \in \{0,1\}$~\cite{Acciai2024Feb}. The ratios of Eq.~\eqref{eq: ratio constant f} are plotted as functions of $f_\text{L}$ for different values of $f_\text{R}$ in Fig.~\ref{fig:constant f}. Panel~(a) shows the result for bosonic systems and panel~(b) the results for fermionic systems. Indeed, both bounds are saturated when one of the reservoirs has a vanishing occupation, while the fermionic one also saturates when one of the occupations is $f_\alpha=1$. The values that the bosonic bounds reach for large occupation differences is calculated by assuming  $f_\text{L}=\delta f+ f_\text{R}$ and taking the limit
\begin{equation}
    \lim_{\delta  f \to \infty} \frac{|f_\text{R}-f_\text{L}|^2}{(F_\text{RL}^++F_\text{LR}^+)^2} = \left(\frac{1}{1+2f_\text{R}}\right)^2.
\end{equation}
This indicates that the key factor influencing the tightness of the bosonic bounds is whether or not one of the reservoirs is empty in the energy interval contributing to transport.

\subsection{Combined temperature and potential bias in bosonic systems}
\begin{figure*}[t]
    \includegraphics[width=6.6in]{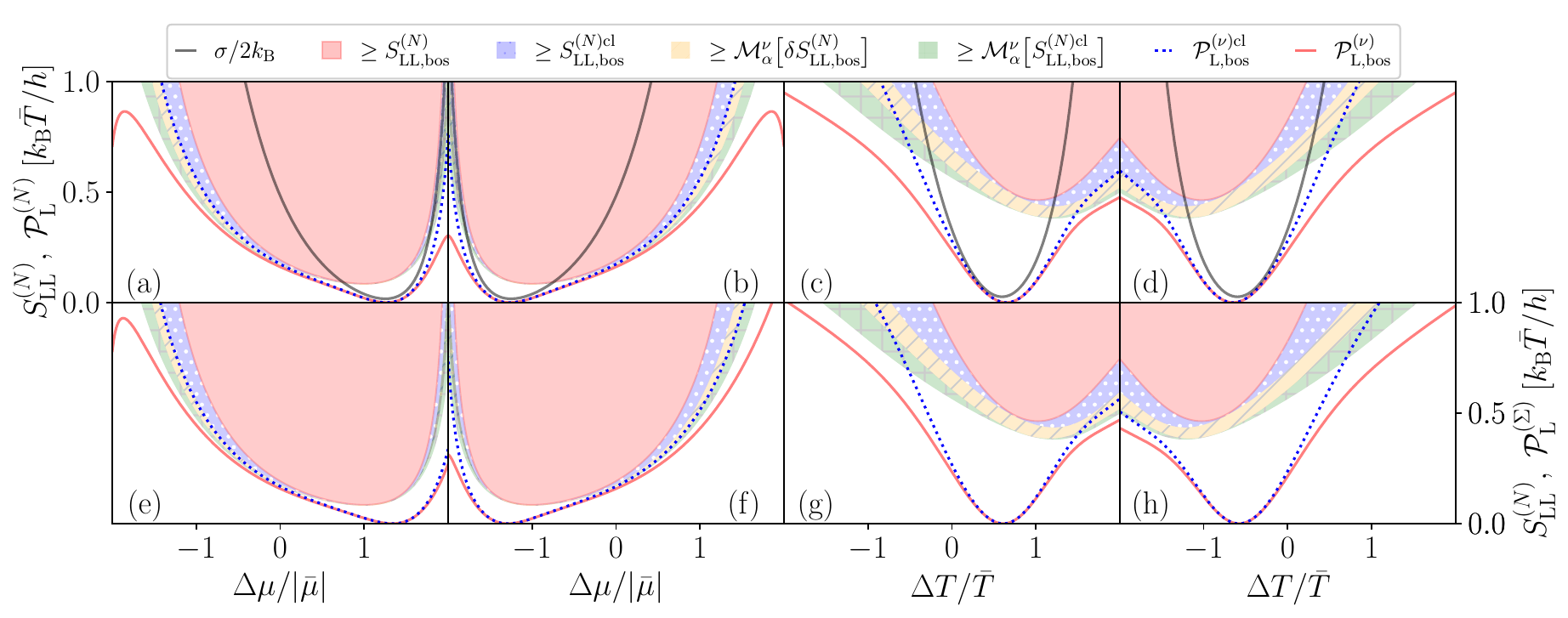}
    \captionsetup{justification=raggedright,singlelinecheck=false, font=small, width=0.9\textwidth} 
\caption{KUR and KURL, see~\eqref{eq:cl-bound}, \eqref{eq:boson-full-weak}, \eqref{eq:half_boson_bound} and~\eqref{eq:full_boson_bound}, in a bosonic two-terminal system.
    Red and blue lines show precision functions, filled areas are the regions excluded by the bounds. Black lines show the total entropy production $\sigma=I^{(\Sigma)}_\text{L}+I^{(\Sigma)}_\text{R}$ constraining precision via the TUR. Upper/lower row: bounds on \textit{particle}-/\textit{entropy}-current precision as function of (a,b,e,f) potential bias, (c,d,g,h) temperature bias. In panels (a,b,e,f) we choose $\bar{\mu}=-3k_\text{B} \bar{T}$ with (a,e) $\Delta T=-0.5\bar{T}$ and (b,f) $\Delta T=0.5\bar{T}$. Panels (c,d,g,h) have $\bar{\mu}=-1.5k_\text{B}\bar{T}$ with (c,g) $\Delta \mu=-0.5|\bar{\mu}|$ and (d,h) $\Delta \mu=0.5|\bar{\mu}|$. For all plots, $b=3k_\text{B} \bar{T}$, $D_0=1$ and $E_0=0.1k_B\bar{T}$.}
    \label{fig:bias plot}
\end{figure*}

In the main text, the presented plots of the precisions and their bounds are shown \textit{either} as function of temperature bias or as function of chemical potential bias, for clarity. It is however interesting from a thermodynamics perspective to examine the bounds when both chemical \textit{and} temperature biases are nonzero, since it is possible to perform work under these boundary conditions. In Fig.~\ref{fig:bias plot}, we hence show equivalent plots to Fig.~\ref{fig:boson_bound} in the main text, but as functions of combined temperature and potential biases. We see that this introduces asymmetries in the dependence on each one of the biases. Concretely, the functions are shifted opposite to the fixed bias along the bottom axis since the biases compete, leading for certain bias pairs to vanishing currents. For this stalling condition, the reservoirs are not at equilibrium, meaning that entropy is still being produced and the TUR is not saturated when the current vanishes. Beyond this ``tilt", the statements of the main text concerning the behavior of precisions and bounds remain valid.

\bibliography{refs.bib}